\documentclass[
 reprint,
superscriptaddress,
 amsmath,amssymb,
 aps,
prb
]{revtex4-2}

\usepackage{pifont}
\usepackage{graphicx}
\usepackage{bm}
\usepackage{sidecap}
\usepackage{xcolor}
\usepackage[colorlinks=true,citecolor=blue]{hyperref}
\usepackage{natbib}
\usepackage{amsmath}
\usepackage{amssymb}
\usepackage{bm}

\makeatletter

\newcommand{\Rmnum}[1]{\expandafter\@slowromancap\romannumeral #1@}
\makeatother

\begin{document}

\title{Tunable high-$Q$ Janus-to-chiral bound states in the continuum in \\bilayer PhCs}

\author{Zhexing Dong}
\affiliation{School of Physics and Optoelectronics, Xiangtan University, Xiangtan 411105, China}

\author{Shengxuan Xia}
\affiliation{School of Physics and Electronics, Hunan University, Changsha 410082, China}
\affiliation{State Key Laboratory of Millimeter Waves, Southeast University, Nanjing 210096, China}

\author{Yee Sin Ang}
\affiliation{Science, Mathematics and Technology (SMT) Cluster, Singapore University of Technology and Design, Singapore 487372, Republic of Singapore}

\author{Haiyu Meng}
\email{h.y.meng@hnu.edu.cn}
\affiliation{School of Physics and Optoelectronics, Xiangtan University, Xiangtan 411105, China}

\begin{abstract}

We propose a bilayer all-dielectric PhC for controlling Janus bound states in the continuum (BIC) and optical chirality through symmetry-selective perturbations. Starting from a symmetry-protected $\Gamma$-point BIC, we use interlayer displacement as one geometric control knob to generate different topological charges in the upward radiation and downward radiation channels. A subsequent diagonal in-plane displacement reconstructs the polarization topology around the BIC and generates a Janus-chiral BIC with strong handedness selectivity. In contrast, other in-plane perturbations generate chiral quasi-BICs with finite radiative coupling, for which the circular dichroism (CD) and resonance wavelength can be continuously tuned. We further show that material conductivity provides an additional dissipative degree of freedom for actively modulating the chiral response, with a switchable CD exceeding 0.89. Near-field optical-chirality distributions and multipole decompositions reveal that the chiral response originates from a symmetry-induced imbalance of local optical handedness and a spin-selective magnetic-dipole resonance. These results reveal the topological relationship between Janus radiation, polarization singularities and intrinsic chirality, thus paving a scalable route toward reconfigurable high-$Q$ chiral photonics.

\end{abstract}
\maketitle

\section{Introduction}

Bound states in the continuum (BICs) have recently emerged as a powerful paradigm for achieving extreme light confinement in open photonic systems, enabling in principle divergent quality factors ($Q$) despite being embedded within the radiation continuum~\cite{Marinica2008PRL,Hsu2016NRM,Kang2023NRP,Wang2024PI,Kang2021PRL,Huang2022PRL,Qin2023LSA,Zhou2023LPR,Sang2024Nanophotonics,Jiang2023PRL,Qin2024NC,Qin2025Nature,Yang2014PRL,Gao2016SciRep,Jin2019Nature,huang2025dynamically}. Photonic crystal (PhC) slabs provide a natural setting for symmetry-protected (SP) and accidental BICs and have enabled lasing~\cite{kodigala2017lasing}, sensing~\cite{yesilkoy2019ultrasensitive,li2019symmetry}, nonlinear-optical processes~\cite{koshelev2020subwavelength,li2019symmetry,liu2025polarization,Ye2022LPR}, and integrated photonic functionalities~\cite{yu2019photonic,Yu2020NatCommun}. From a topological viewpoint, BICs correspond to polarization vortices in momentum space carrying integer topological charges. This charge structure underlies their robustness against continuous perturbations and provides a versatile framework for engineering far-field radiation~\cite{zhen2014topological,Jin2019Nature,Liu2019PRL,Wang2020NatPhotonics,Wang2022PRL,Hu2022Optica,Jiang2023PRL,Hu2023NSR,Wu2024PhotonicsResearch,ye2020singular}.

Recent efforts have sought to extend this topological control toward directional and chiral photonic responses~\cite{Kang2022NatCommun,Qin2023LSA,Shi2022NatCommun,Lv2024SciAdv,Tu2024PhotonicsResearch}. Janus BICs, defined by inequivalent topological charges in the upward and downward radiation channels, lift the symmetry constraints that ordinarily forbid the coexistence of intrinsic chirality with diverging $Q$~\cite{ji2025janus,kang2025janus,yin2025janus,zuo2025janus}. Chiral BICs and chiral quasi-BICs, in turn, have emerged as platforms for strong circular dichroism (CD) and enhanced chiral light-matter interactions~\cite{gorkunov2020metasurfaces,chen2023observation,chen2021chirality,Shi2022NatCommun,Barkaoui2023PRB,Duan2023PhotonicsResearch,Wan2022OptLett,Tang2023LPR,Lv2024SciAdv,Tu2024PhotonicsResearch,yi2025efficient,cao2025double,tong2025nearly,qin2024rotation,zhao2023resonance}. In most existing realizations, however, intrinsic chirality is generated by symmetry breaking that converts an ideal BIC into a leaky quasi-BIC, imposing an intrinsic trade-off between achievable CD and $Q$~\cite{gorkunov2020metasurfaces,chen2021chirality,Shi2022NatCommun,Barkaoui2023PRB,Duan2023PhotonicsResearch,Wan2022OptLett,Tang2023LPR,zhao2024spin,zhou2024chiral,mo2025brillouin,kim2024tunable,yang2025phase}.

  \begin{figure*}
    \centering
    \includegraphics[width=1\linewidth]{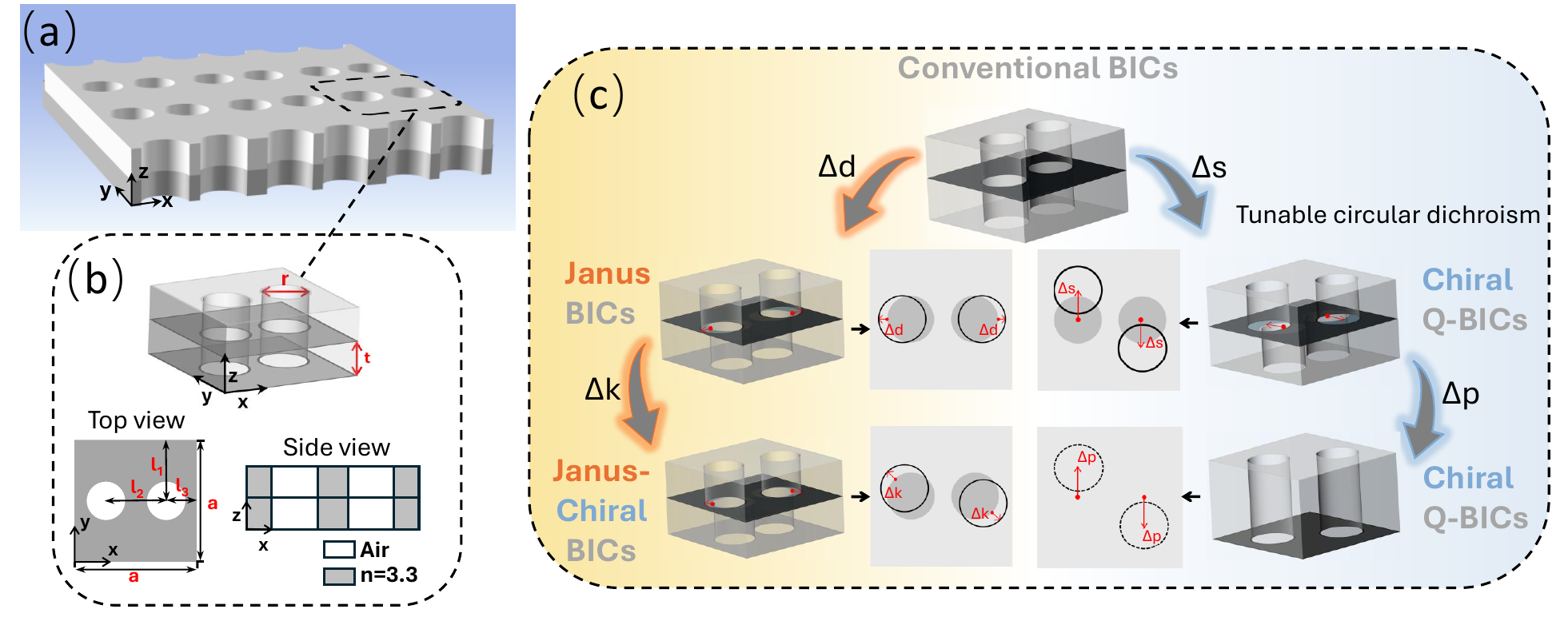}
    \caption{Concept and model structures of the bilayer PhC hosting tunable Janus chiral BICs. (a) Three-dimensional (3D) view of the two-dimensional (2D) periodic air-hole array. The gray regions represent the host material with a refractive index of $n=3.3$, while the holes are filled with air. (b) Enlarged views of the single unit cell indicated by the dashed box, including 3D, top, and side views. The key geometric parameters are defined as follows: lattice constant $a = 481$~nm, slab thickness $t = 136$~nm, and air hole radius $r = 80$~nm. The center positions of the holes are determined by the parameters $l_1 = 0.5a$, $l_2 = 202$~nm, and $l_3 = 139.5$~nm. (c) Hierarchical control map summarizing the distinct roles of different control knobs in enabling Janus and chiral bound states in the continuum. The parameters $\Delta d$, $\Delta s$, and $\Delta k$ enable both Janus and chiral BICs, while the collective in-plane displacement $\Delta p$ and material loss $\sigma$ provide additional control of chiral BICs only.}
\label{fig:structure}
\end{figure*}

Functional applications such as on-chip chiral light sources~\cite{huang2020ultrafast,zhou2015chip,lodahl2017chiral,yang2021perspective,he2014chip,kim2021chiral}, reconfigurable CD filters~\cite{yin2022active,jiang2023shape,zhou2020switchable,jiang2023temperature,yin2018reconfigurable}, chiral-transmission devices~\cite{shu2024chiral}, and polarization-encoded optical switches~\cite{yang2019nonlinear,zeng2025reconfigurable,ding2021recent,mandal2017design,samanta2012all,raja2021analysis} require programmable control of chirality magnitude, operating frequency and radiation directionality, together with dynamic switching, within a single device architecture. Recent studies have shown that Janus radiation topology and chiral responses can be closely connected~\cite{ji2025janus,kang2025janus,yin2025janus,zuo2025janus}. Nevertheless, most existing schemes rely on a limited set of symmetry-breaking perturbations, making it difficult to separately control Janus directionality, intrinsic chirality, spectral position, and active switching within a unified framework. A systematic understanding of how different symmetry channels affect these optical responses is therefore still needed for  controllable high-$Q$ chiral photonics. 

In this work, we introduce a bilayer all-dielectric PhC platform that supports Janus BICs, Janus-chiral BICs, and tunable chiral quasi-BIC responses through distinct symmetry-breaking pathways. Starting from a SP-BIC, an interlayer displacement decouples the upward and downward radiation channels and produces a Janus BIC, providing the topological intermediate from which a Janus-chiral BIC is obtained by an additional diagonal in-plane displacement, with the divergent $\Gamma$-point $Q$ preserved along this Janus-to-chiral evolution pathway. Independent in-plane perturbations instead drive the system into chiral quasi-BICs whose CD amplitude and resonance wavelength are continuously tunable, and a tunable bulk conductivity applied to the fixed antiphase-displaced configuration enables active and reversible modulation of the chiral response without further geometric reconfiguration. These results provide a symmetry-based description of how Janus radiation, polarization-singularity evolution, and optical chirality can be controlled by distinct perturbation channels in bilayer PhCs.

\section{Structure and simulation model}

Figure~1 shows the schematic of the bilayer PhC slab considered in this work. 
The structure consists of two vertically stacked all-dielectric slabs with refractive index $n=3.3$, perforated by cylindrical air holes arranged in a square lattice. The lattice constant is $a=481$~nm, the thickness of each slab is $t=136$~nm, and the air-hole radius is $r=80$~nm. To maintain a vertically symmetric reference configuration, identical substrate and superstrate layers with refractive index $n=1.5$ are placed below and above the bilayer slab. In each unit cell, two air holes are patterned in each layer. Their unperturbed positions are determined by the geometric parameters $l_1=0.5a$, $l_2=202$~nm, and $l_3=139.5$~nm. 

The central idea of this work is to use the bilayer geometry to access different symmetry-breaking pathways within the same photonic platform. Instead of relying on a single perturbation, we introduce several displacement parameters that modify different spatial symmetries and therefore control different far-field radiation topology and polarization-selective chirality. The interlayer displacement $\Delta d$ shifts the air holes in the upper layer relative to those in the lower layer and breaks the out-of-plane mirror symmetry $\sigma_z$. This perturbation lifts the symmetry-imposed equivalence between the upward and downward radiation channels and provides a route to Janus BICs. By contrast, the orthogonal displacement $\Delta s$ and the collective displacement $\Delta p$ represent independent in-plane control knobs for tuning the chiral response, forming chiral quasi-BIC responses. These perturbations do not necessarily rely on the Janus radiation topology, but provide additional routes to engineer CD and its spectral position. Another displacement, denoted by $\Delta k$, is applied along the diagonal direction. When an in-plane perturbation is applied to a Janus configuration, these two effects coexist and give rise to a Janus-chiral BIC. Finally, a tunable material conductivity $\sigma$ is introduced as a dissipative control parameter for active modulation of the chiral response.

    The optical properties are calculated using the finite-element method implemented in COMSOL Multiphysics. We solve the full-vectorial Maxwell eigenvalue problem in the frequency domain. Bloch periodic boundary conditions are applied along the in-plane $x$ and $y$ directions, while perfectly matched layers are used along the out-of-plane $z$ direction to simulate an open radiation environment. The resonant frequencies and Q factors are determined from the real and imaginary parts of the complex eigenfrequencies, while the electromagnetic field distributions are extracted from the corresponding complex eigenmodes. For transmission and CD calculations, normally incident left-handed circularly polarized (LCP) and right-handed circularly polarized (RCP) plane waves are used to obtain the polarization-resolved transmission coefficients. These quantities form the foundation for systematically analyzing the impact of different symmetry-breaking pathways on the emergence, evolution, and controllability of Janus and chiral BICs.

 \begin{figure*}
    \centering
    \includegraphics[width=1\linewidth]{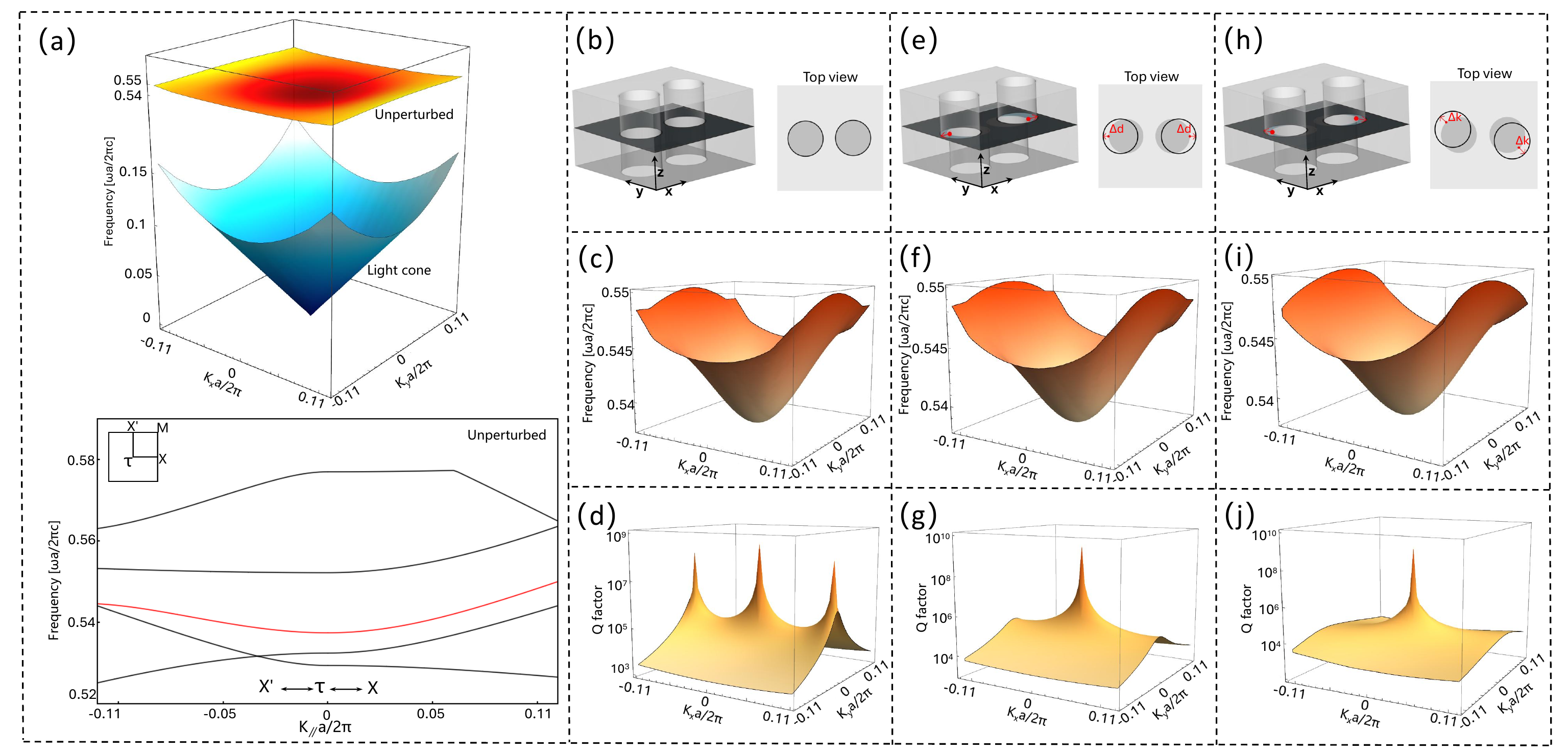}
    \caption{Lattice structure, band structure and radiative $Q$-factor distribution of the bilayer PhCs. (a) shows the band structure, where the blue surface indicates the light cone boundary. (b) Symmetric configuration preserving $\sigma_z$ symmetry, (c) the corresponding band structure, and (d) the radiative $Q$-factor distribution in momentum space. Both the SP-BIC at the $\Gamma$ point and the accidental BICs at the off-$\Gamma$ points exhibit ultra-high $Q$-factors. (e)--(g) and (h)--(j) show the same quantities as (b)--(d), but for two asymmetric configurations in which the upper air holes are displaced in opposite directions along the $x$-axis and along the $45^{\circ}$ diagonal direction, respectively. In both asymmetric configurations, the BIC at the $\Gamma$ point is robust, maintaining an ultra-high $Q$-factor. Conversely, the accidental BICs at the off-$\Gamma$ points undergo selective destabilization and degenerate into leaky modes due to the breaking of $\sigma_z$ symmetry.}
\label{fig:band_evolution}
\end{figure*}

\section{Tunable Janus-chiral BIC platform and functional control}

\subsection{SP-BIC as the starting point of the platform}

In Fig. 2(a), we first provide the eigenmode spectrum of the unperturbed bilayer PhC, whose overall structure and top view are presented in Fig.~2(b). The upper panel in Fig. 2(a) shows the 3D dispersion in momentum space, with the blue surface denoting the light cone and the red-yellow surface positioned above it. The lower panel provides an enlarged one-dimensional band structure within the frequency range of $0.52$--$0.59$ along the high-symmetry path. This band lies inside the radiation continuum and is therefore, in principle, allowed to couple to outgoing plane waves. This configuration preserves the out-of-plane mirror symmetry $\sigma_z$, as the air holes in the two layers are vertically aligned [Fig.~2(b)]. The corresponding dispersion and radiative $Q$-factor distribution are shown in Figs.~2(c) and 2(d). Two types of BICs can be identified. The first one is located at the $\Gamma$ point and corresponds to a SP-BIC. Its vanishing radiation originates from the incompatibility between the symmetry of the eigenmode and the outgoing plane-wave radiation channel at normal incidence. The second one appears away from the $\Gamma$ point and corresponds to an accidental BIC, which is formed by destructive interference between different radiation pathways.

To investigate the response of these BICs to symmetry breaking, we introduce interlayer displacements of the upper air holes relative to the lower layer, either along the $x$ axis [Fig.~2(e)] or along the $45^{\circ}$ diagonal direction [Fig.~2(h)], thereby explicitly breaking the $\sigma_z$ symmetry. A comparison of the resulting dispersion surfaces in Figs.~2(f) and 2(i) with the unperturbed case in Fig.~2(c) indicates that the eigenfrequencies remain remarkably robust against these structural perturbations. As shown in Figs.~2(b), 2(e), and 2(h), the applied interlayer displacements modify only the relative in-plane positions of the air holes while preserving the overall slab thickness, material composition, and lattice periodicity. Consequently, the effective mode and vertical confinement of the resonant modes remain nearly unchanged. This is directly reflected in the almost identical dispersion surfaces observed in Figs.~2(c), 2(f), and 2(i), indicating that the real parts of the eigenfrequencies are remarkably insensitive to the applied structural perturbations. In contrast, the interlayer displacement breaks the out-of-plane mirror symmetry and therefore strongly alters the interference conditions of the outgoing radiation channels, leading to pronounced modifications of the radiative losses without significantly affecting the modal frequencies. Additionally, the corresponding Q-factor distributions [Figs.~2(g) and 2(j)] can also demonstrate this behavior, which exhibit a pronounced and highly selective stabilization behavior. While the $\Gamma$-point SP-BIC remains robust and preserves an ultrahigh $Q$ factor, the accidental BIC undergoes a rapid collapse in $Q$ and transform into leaky resonances.

This contrasting behavior directly reflects their distinct formation mechanisms. The accidental BICs critically rely on out-of-plane mirror symmetry to maintain destructive interference between radiation channels. Once $\sigma_z$ is broken, this interference condition is violated, leading to inevitable radiation leakage. By contrast, the SP-BIC is protected by in-plane symmetry and is therefore insensitive to vertical asymmetry. This mechanism enables the selective elimination of the accidental BICs while preserving the $\Gamma$-point SP-BIC, creating a pronounced asymmetry between radiation channels. As will be shown in the following, this selective response provides the physical basis for the Janus BIC discussed in the next section. By breaking $\sigma_z$, the two radiation channels are no longer constrained to carry identical polarization singularity structures. Consequently, the topological charge evolution in the upward and downward channels can become asymmetric while the $\Gamma$-point BIC remains intact.

\subsection{Janus BICs from radiation-channel decoupling}

\begin{figure*}[htbp]
\centering
\includegraphics[width=1\linewidth]{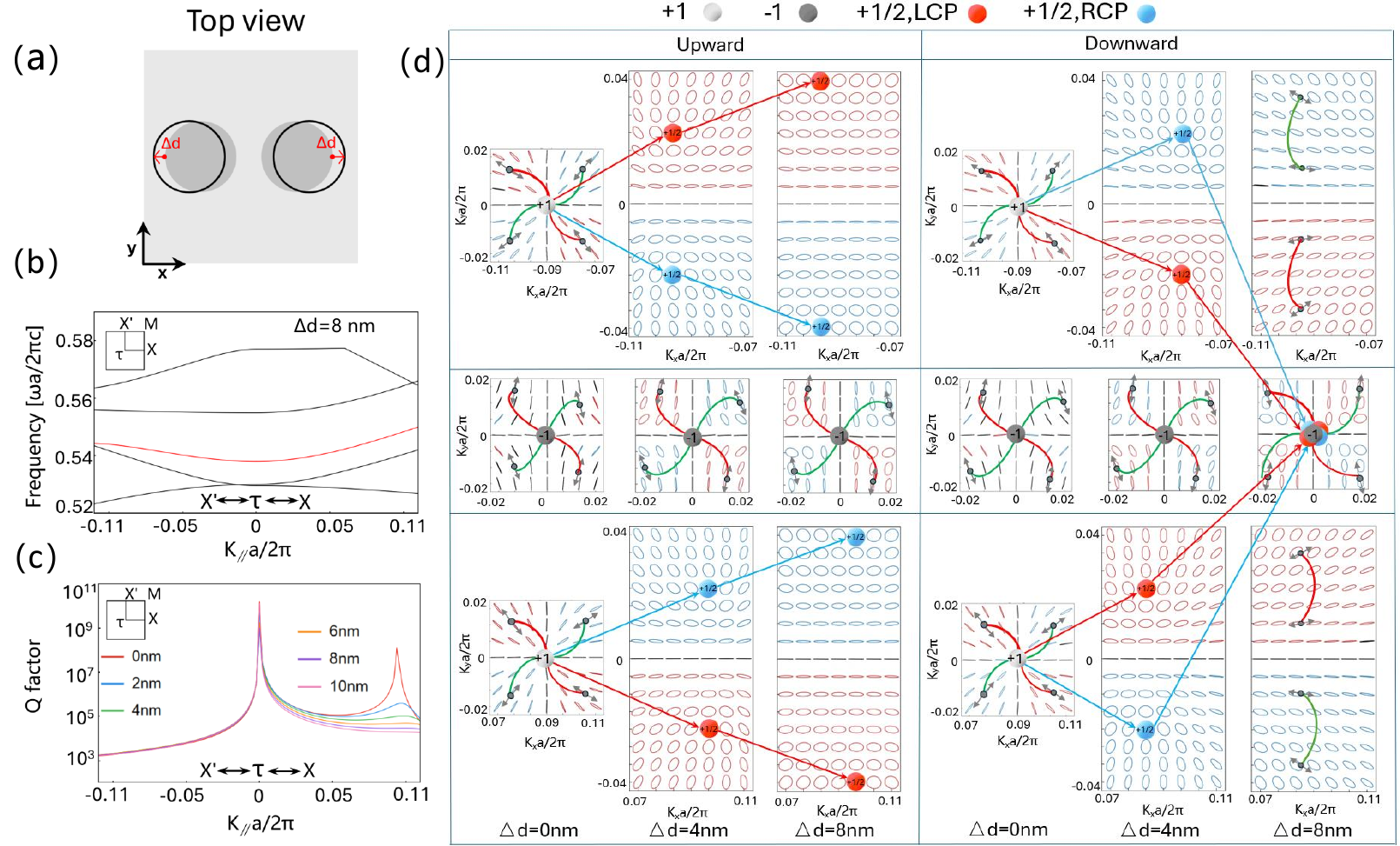}
\caption{
Janus BIC induced by radiation-channel decoupling.
\textbf{(a)} Top view of the unit cell with an interlayer displacement $\Delta d$, where the upper-layer air holes are shifted along the $x$ direction relative to the lower layer, breaking the out-of-plane mirror symmetry $\sigma_z$ while preserving the in-plane symmetry protecting the $\Gamma$-point BIC.
\textbf{(b)} Band structure at $\Delta d = 8~\mathrm{nm}$, showing the relevant $\Gamma$-point mode.
\textbf{(c)} Evolution of the radiative $Q$ factor with $\Delta d$. Upon $\sigma_z$ breaking, the off-$\Gamma$ accidental BICs rapidly lose their divergent $Q$ factors, whereas the $\Gamma$-point SP-BIC remains robust with an ultrahigh $Q$ factor.
\textbf{(d)} Evolution of far-field polarization singularities in the upward and downward radiation channels. At $\Delta d=0$, both channels exhibit identical polarization-vortex configurations, with a central $q=-1$ vortex at $\Gamma$ and off-$\Gamma$ accidental BICs carrying $q=+1$. For $\Delta d>0$, the off-$\Gamma$ integer vortices split into C points and evolve asymmetrically in the two radiation channels, leading to different net topological charges enclosed around $\Gamma$. This channel-dependent topological evolution reveals the formation of a Janus BIC. White and gray dots denote integer charges of $+1$ and $-1$, respectively, while red and blue dots denote LCP and RCP C points with half-integer charges, respectively.
}
\end{figure*}

We next introduce controlled breaking of the out-of-plane mirror symmetry to selectively manipulate radiation channels, thereby enabling the formation of Janus BICs. As shown in Fig.~3(a), an in-plane displacement $\Delta d$ is introduced to break the $\sigma_z$ symmetry. The band structure for $\Delta d = 8~\mathrm{nm}$ is shown in Fig.~3(b). Fig.~3(c) illustrates the evolution of $Q$-factor with increasing $\Delta d$. As $\sigma_z$ symmetry is lifted, the accidental off-$\Gamma$ BICs, which are sustained by destructive interference, evolve into leaky resonances, accompanied by a significant reduction in the $Q$-factor. In contrast, the BIC at the $\Gamma$ point, protected by the in-plane $C_2$ rotational symmetry, exhibits remarkable robustness against this specific geometric perturbation, maintaining an ultrahigh $Q$-factor. This differentiated radiation behavior provides the necessary physical environment for the formation of the Janus state.

To elucidate the topological origin of the Janus BIC, we analyze the evolution of polarization singularities in momentum space. The topological charge $q$ is defined as the winding number of the polarization azimuth angle $\phi(\mathbf{k})$ along a closed loop $l$ surrounding the singularity\cite{zhen2014topological}: 
\begin{equation}
    q = \frac{1}{2\pi} \oint_{l} d\mathbf{k} \cdot \nabla_{\mathbf{k}} \phi(\mathbf{k})
    \label{eq:topological_charge}
\end{equation}
where $\phi(\mathbf{k})$ describes the orientation of the polarization state in momentum space. Following previous formulations~\cite{zhen2014topological,ye2020singular}, $\phi(\mathbf{k})$ is defined by the argument of the complex Fourier coefficients:
\begin{equation}
    \phi(\mathbf{k}) = \arg[C_x(\mathbf{k}) + i C_y(\mathbf{k})]
    \label{eq:polarization_angle}
\end{equation}
Here, $C_x(\mathbf{k})$ and $C_y(\mathbf{k})$ represent the Fourier coefficients of the electric field components, which characterize the spatial frequency content of the electromagnetic field in the periodic structure\cite{ye2020singular}:
\begin{align}
    C_x(\mathbf{k}) &= \frac{1}{\iint_{\text{cell}} dx dy} \iint_{\text{cell}} E_x^* e^{-i(k_x x + k_y y)} dx dy \label{eq:Cx_def} \\
    C_y(\mathbf{k}) &= \frac{1}{\iint_{\text{cell}} dx dy} \iint_{\text{cell}} E_y^* e^{-i(k_x x + k_y y)} dx dy \label{eq:Cy_def}
\end{align}
where the integration is performed over the unit cell area, and $E_{x,y}^*$ denotes the complex conjugate of the electric field components.

As illustrated in Fig.~3(d), in the symmetric configuration ($\Delta d=0$~nm), the accidental BICs at the off-$\Gamma$ points manifest as polarization vortices carrying integer topological charges of $q=+1$ (V-points, light gray dots) in momentum space, while the SP-BIC at the $\Gamma$ point carries a topological charge of $q=-1$ (dark gray dot). At this stage, both the upward and downward radiation channels exhibit identical topological features. Once the out-of-plane asymmetry is introduced ($\Delta d > 0$), the off-$\Gamma$ integer topological charges undergo splitting, each degenerating into a pair of circularly polarized points (C-points) carrying half-integer topological charges of $q=+1/2$ (denoted by red and blue dots for LCP and RCP, respectively). 

Notably, the absence of $\sigma_z$ symmetry leads to a significant bifurcation in the migration trajectories of these C-points in momentum space, as explicitly traced by the solid arrows. In the downward radiation channel, the split C-points migrate inward and eventually undergo topological recombination at the $\Gamma$ point when $\Delta d = 8$~nm, forming a net topological charge characteristic of $q=+1$. Conversely, in the upward radiation channel, the C-points are repelled outward, leaving the topological configuration at the $\Gamma$ point robustly pinned at $q=-1$. This decoupled topological evolution, where orthogonal radiation channels carry asymmetric net topological charges within the same resonant mode, unambiguously marks the successful realization of the Janus BIC.

\subsection{Janus-chiral BIC generated by combined interlayer and in-plane perturbations}

Having established the Janus BIC through interlayer displacement, we next fix this Janus configuration and introduce intrinsic chirality by applying an additional diagonal in-plane displacement parameter $\Delta k$, as shown in Fig.~4(a). In this section, $\Delta k$ is the only tuning parameter. Unlike the interlayer displacement $\Delta d$, which has already been used to break the out-of-plane mirror symmetry $\sigma_z$ and decouple the upward and downward radiation channels, the diagonal displacement $\Delta k$ further breaks the remaining in-plane mirror symmetries. Importantly, the rotational symmetry required to protect the $\Gamma$-point BIC is preserved. Therefore, the additional $\Delta k$ perturbation provides a route to a Janus-chiral BIC, where the previously established Janus radiation topology and the newly induced handedness-selective polarization topology coexist around the same high-$Q$ resonance.

As confirmed by the $Q$-factor distribution in Fig.~4(b), the $\Gamma$-point mode remains a true BIC with divergent $Q$ after introducing $\Delta k=10~\mathrm{nm}$. This indicates that the emergence of chirality does not rely on converting the BIC into a leaky quasi-BIC. Instead, the diagonal displacement $\Delta k$ reconstructs the polarization topology around the BIC singularity while maintaining the nonradiative BIC condition at normal incidence.

The origin of the chiral polarization topology is elucidated by the different behavior of C points in momentum space, as shown in Figs.~4(c) and 4(d). Driven by the $\Delta k$ perturbation, the upward radiation channel exhibits a selective C-point coalescence process: only the C points carrying LCP are tuned to the $\Gamma$ point, while the RCP C points remain separated from $\Gamma$. In contrast, the C-point configuration in the downward radiation channel follows a different evolution. Consequently, the upward and downward radiation channels acquire distinct polarization-singularity structures and different effective topological charge configurations, with the Janus BIC carrying $q^u=0$ and $q^d=+1$. This channel-dependent recombination of C points is the microscopic origin of the Janus-chiral polarization response.

To quantitatively characterize the optical chirality associated with this topological state, we calculate the momentum-space distribution of the normalized Stokes parameter $S_3$, which measures the degree of circular polarization in the far-field radiation:
\begin{equation}
    S_3(\mathbf{k})=
    \frac{2\,\mathrm{Im}(E_xE_y^*)}{|E_x|^2+|E_y|^2}
    \label{eq:stokes_s3}
\end{equation}
where $E_x$ and $E_y$ are the far-field electric-field components in the selected radiation channel.

The momentum-space polarization distributions and the corresponding $S_3$ maps for the upward and downward radiation are shown in Figs.~4(c) and 4(d). Locally, a prominent feature is the emergence of near-unity circular polarization in the immediate vicinity of the $\Gamma$ point, as indicated by the white contour lines of $|S_3|\ge 0.95$. The Poincar\'e-sphere projections in Figs.~4(c) and 4(d) further reveal a channel-dependent reconstruction of the far-field polarization topology. The upward and downward radiation channels occupy distinct regions on the Poincar\'e sphere, indicating that their polarization states are no longer related by mirror symmetry. This difference is consistent with the asymmetric evolution and topological recombination of $C$ points in the two radiation channels, and leads to opposite circular-polarization tendencies in the two propagation directions.

To further resolve the singular behavior near the $\Gamma$ point, Fig.~4(e) plots the line cut of the normalized Stokes parameter $S_3$ along the $k_x$ direction. Near the Brillouin-zone center, the upward and downward radiation channels exhibit opposite circular-polarization tendencies, as evidenced by the opposite signs of $S_3$. However, the value of $S_3$ exactly at $\Gamma$ should not be interpreted as a conventional far-field polarization response. Since the central mode remains a true BIC at normal incidence, its far-field radiation amplitude vanishes at the $\Gamma$ point. Consequently, the normalized Stokes parameter is physically ill-defined exactly at the BIC singularity, although the polarization states in its immediate momentum-space neighborhood can approach nearly pure circular polarization.

It should be emphasized that Figs.~4(a)--4(e) mainly reveal the Janus-chiral polarization topology around the $\Gamma$-point BIC, rather than demonstrating a finite CD exactly at normal incidence. At the exact true-BIC point, the far-field radiation amplitudes vanish, so the polarization state and the corresponding LCP/RCP scattering response are not well defined. Therefore, CD cannot be strictly assigned to the BIC itself. Instead, the chiral character of the BIC should be inferred from the limiting far-field response in the vicinity of the BIC or from a symmetry-perturbed quasi-BIC with finite radiative coupling. Alternatively, a measurable high-CD response requires a finite in-plane wave vector, corresponding to near-normal oblique incidence. This interpretation is consistent with the analysis by Kang \textit{et al.}~\cite{kang2025janus}. To obtain a well-defined circularly polarized far-field response strictly at normal incidence, additional symmetry perturbations can be introduced to reduce the $Q$ factor and transform the state into a chiral quasi-BIC. In contrast, our design prioritizes the preservation of the true BIC and its ultrahigh-$Q$ confinement at normal incidence. Consequently, strong chirality is accessed in the finite-$k$ momentum-space region immediately surrounding the BIC singularity, providing a practical trade-off between high spectral selectivity and strong chiral response.

Furthermore, the evolution of the CD spectrum with $\Delta k$ provides a comprehensive view of the tunability of the finite-$k$ chiral response, as shown in Fig.~4(f). Here, the CD is evaluated in the momentum-space vicinity of the $\Gamma$-point BIC rather than exactly at the BIC singularity. At small $\Delta k$, the surrounding radiative states have not yet acquired significant handedness selectivity, and the system remains nearly achiral. As $\Delta k$ increases, a high-CD branch emerges and becomes strongly enhanced, reaching a maximum value exceeding 0.8 around $\Delta k\approx 30~\mathrm{nm}$. Meanwhile, the resonance wavelength exhibits a blueshift from approximately $968~\mathrm{nm}$ to $956~\mathrm{nm}$. These results indicate that the diagonal displacement $\Delta k$ acts as an effective tuning parameter for both the operating wavelength and the strength of the finite-$k$ chiral response, while the exact $\Gamma$-point state remains a nonradiative high-$Q$ BIC.

\begin{figure*}
    \centering
    \includegraphics[width=1\linewidth]{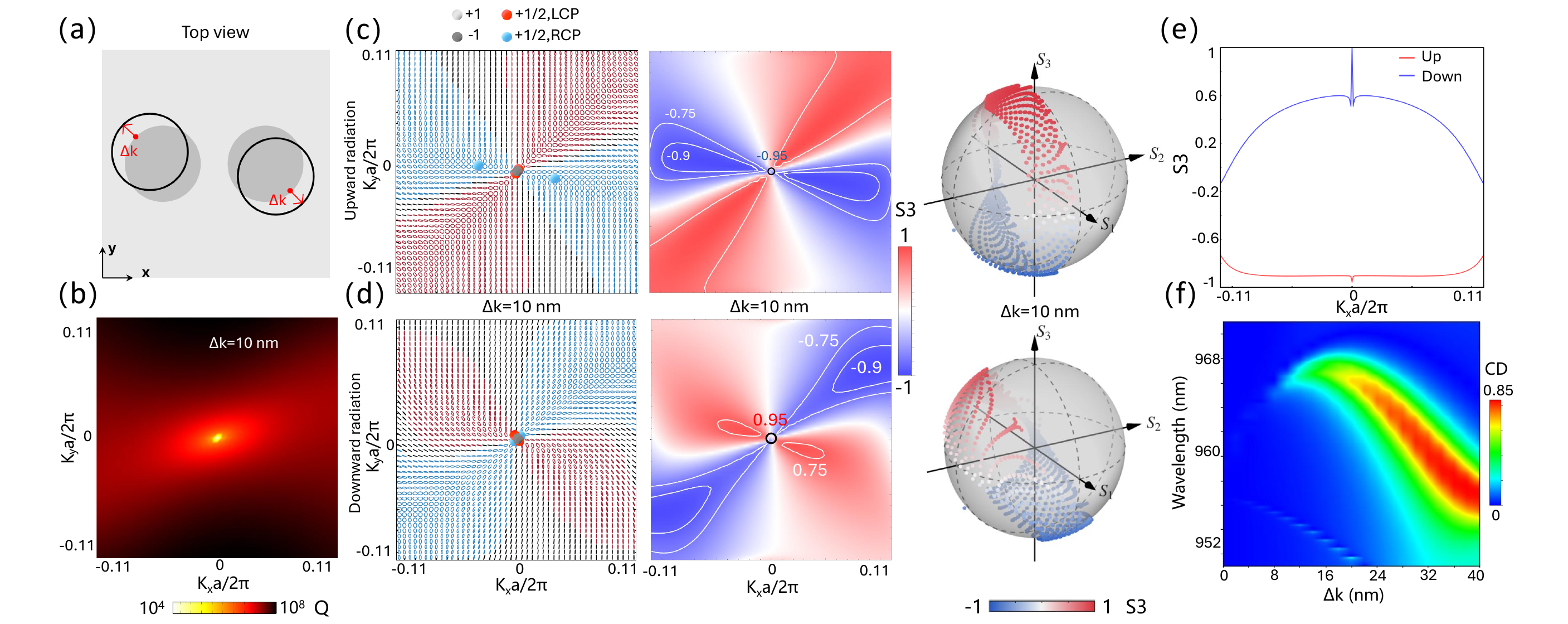}
    \caption{Intrinsic chiral Janus BICs enabled by in-plane symmetry breaking.
    \textbf{(a)} Top view of the unit cell with a diagonal displacement perturbation $\Delta k$ introduced to break the in-plane mirror symmetry.
    \textbf{(b)} Radiative $Q$-factor distribution in momentum space at $\Delta k = 10$~nm, confirming that the $\Gamma$-point mode remains a high-$Q$ BIC.
    \textbf{(c)} Polarization topology of the upward radiation channel at $\Delta k = 10$~nm, including the polarization-ellipse map, the normalized Stokes parameter $S_3$ distribution, and the corresponding Poincaré-sphere projection. Red and blue dots denote LCP and RCP C-points with half-integer topological charges, respectively, while gray dots denote integer-charge polarization singularities. The upward channel exhibits selective coalescence of LCP C-points near the $\Gamma$ point, giving rise to an effective topological charge of $q^{u}=0$.
    \textbf{(d)} Polarization topology of the downward radiation channel at $\Delta k = 10$~nm. Compared with the upward channel, the downward channel shows a distinct C-point recombination process and an opposite circular-polarization tendency around the $\Gamma$ point, resulting in an effective topological charge of $q^{d}=+1$.
    \textbf{(e)} Line profiles of the normalized Stokes parameter $S_3$ along the $k_x$ direction for the upward and downward radiation channels, showing opposite circular-polarization responses near the $\Gamma$ point.
    \textbf{(f)} Evolution of the CD spectrum as a function of the perturbation parameter $\Delta k$, showing the emergence and enhancement of the finite-$k$ chiral response.}
\end{figure*}

To quantitatively characterize this chiral optical response, CD is defined as the differential transmission between LCP and RCP light:
\begin{equation}
\begin{aligned}
    \mathrm{CD} &= |t_{LL}|^2 + |t_{RL}|^2 - \left(|t_{RR}|^2 + |t_{LR}|^2\right) \\
                &= T_{\mathrm{LCP}} - T_{\mathrm{RCP}} 
\end{aligned}
\label{eq:cd_definition}
\end{equation}
Here, $t_{ij}$ ($i,j \in \{L,R\}$) denotes the transmission coefficient from incident polarization state $j$ to transmitted polarization state $i$. Accordingly, $T_{\mathrm{LCP}} = |t_{LL}|^2 + |t_{RL}|^2$ represents the total power transmission under LCP incidence, while $T_{\mathrm{RCP}} = |t_{RR}|^2 + |t_{LR}|^2$ corresponds to that under RCP incidence. This definition accounts for polarization conversion, thereby avoiding an artificial overestimation of CD based solely on co-polarized transmission. The simultaneous realization of high CD and an ultra-high $Q$ factor highlights the potential of this design for enhanced chiral light--matter interactions.

\subsection{Independent geometric tuning of chiral responses by orthogonal \textit{y}-axis shift}

We next show that the chiral response of the bilayer PhC slab can also be controlled 
through an independent in-plane displacement pathway. 
In contrast to the combined perturbation $(\Delta d,\Delta k)$ discussed above, here we introduce 
an orthogonal displacement $\Delta s$ along the $y$ direction, as illustrated in Fig.~5(a). This perturbation modifies the in-plane mirror symmetries and changes the handedness-dependent coupling of the resonant mode, while it does not rely on a preexisting Janus BIC. Therefore, $\Delta s$ provides an independent geometric degree of freedom for tuning the circular-polarization response in the same bilayer platform. Figure~5(b) shows the evolution of the CD spectra as a function of $\Delta s$. For small displacement, the structure remains nearly achiral and the transmission spectra for LCP and RCP light are almost degenerate. 
As $\Delta s$ increases, the degeneracy between the two circular-polarization channels is lifted, and a pronounced CD signal gradually develops. The maximum CD exceeds 0.8 over a broad range of displacement, demonstrating that the chiral response can be continuously enhanced by geometric tuning. At the same time, the resonance wavelength shifts from approximately 962~nm to 948~nm, indicating that $\Delta s$ simultaneously controls the strength and spectral position of the chiral response.

The polarization origin of this tunability is further clarified by the evolution of the normalized Stokes 
parameter $S_3$, shown in Fig.~5(c). The results reveal that as $\Delta s$ increases from 0 to 80~nm, the $S_3$ parameters for the upward and downward radiation channels undergo a significant separation process, asymptotically approaching perfect LCP ($S_3 \to 1$) and RCP states ($S_3 \to -1$), respectively. This progressive evolution of polarization states quantitatively explains the substantial enhancement of the macroscopic CD value driven by the increased structural asymmetry. Notably, at $\Delta s = 80$~nm [Fig.~5(d)], the CD spectrum exhibits a sharp sign reversal and high amplitude characteristics near 963~nm. This drastic polarization transition occurring within a narrow spectral range renders the structure highly suitable for applications in high-sensitivity chiral switches, polarization modulators, and chiral biosensors.

\begin{figure}
    \centering
    \includegraphics[width=1\linewidth]{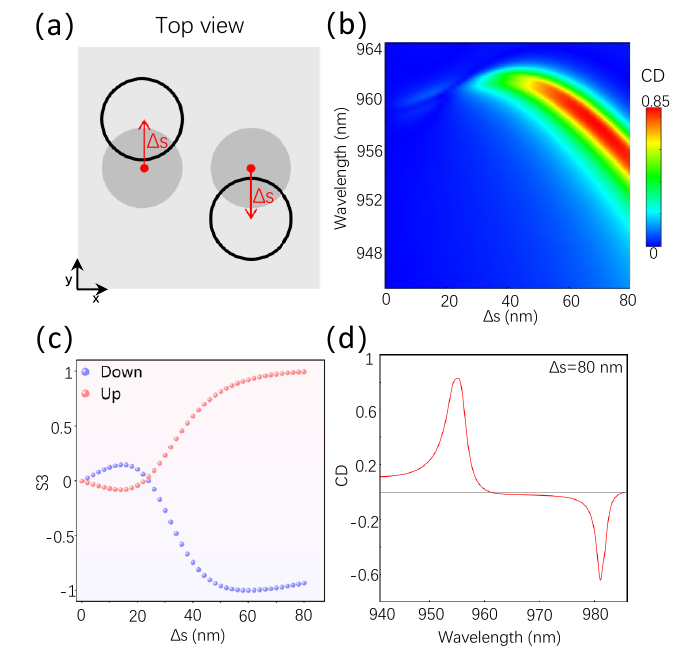}
    \caption{Tunable chirality and switching functionality enabled by structural parameter $\Delta s$. 
\textbf{(a)} Top view of the unit cell with an opposing $y$-axis displacement $\Delta s$. 
\textbf{(b)} Evolution of CD spectra as a function of $\Delta s$. 
\textbf{(c)} Evolution curves of the normalized Stokes parameter $S_3$ for upward (red line) and downward (blue line) radiation channels versus $\Delta s$. As $\Delta s$ increases, the $S_3$ values diverge and asymptotically approach $+1$ and $-1$, respectively. 
\textbf{(d)} CD spectrum at $\Delta s = 80$~nm, exhibiting a drastic polarization transition near 963~nm.}
\label{fig:tunable_chirality}
\end{figure}

To unveil the physical origin of CD, we analyze the near-field optical chirality density (OCD) of the resonant modes. Following the definition used in Ref.~\cite{Li2025PermittivityAsymmetric}, the OCD is evaluated as:
\begin{equation}
    \eta_{\mathrm{OCD}} 
    = -\frac{\omega}{2} \operatorname{Re}\left[\mathbf{D} \cdot \mathbf{B}^{*}\right]
    = -\frac{\omega \varepsilon_0 \varepsilon_r}{2}
    \operatorname{Re}\left[\mathbf{E} \cdot \mathbf{B}^{*}\right]
    \label{eq:ocd_definition}
\end{equation}
where $\omega$ is the angular frequency, and $\mathbf{E}$, $\mathbf{D}$, and $\mathbf{B}$ denote the complex amplitudes of the electric field, electric displacement field, and magnetic induction, respectively. For an isotropic dielectric medium, $\mathbf{D}=\varepsilon_0\varepsilon_r\mathbf{E}$.

In the unperturbed symmetric configuration (left column, $\Delta s = 0$~nm), the system exhibits no intrinsic chirality. As depicted in Fig.~6(a), the OCD distribution displays a strict antisymmetric character, where regions of positive (red) and negative (blue) chirality density alternate with equal magnitude. This antisymmetric distribution leads to a vanishing net optical chirality after spatial integration, consistent with the absence of circular-polarization selectivity in the far field, which is further corroborated by the linear polarization maps for both upward [Fig.~6(b)] and downward [Fig.~6(c)] radiation, and explicitly confirmed by the degenerate transmission spectra in Fig.~6(d).

In stark contrast, the right column illustrates the system response under a broken-symmetry configuration with $\Delta s = 70$~nm. The geometric misalignment disrupts the equilibrium of the near-field OCD [Fig.~6(e)], creating an uncompensated net chiral component. This near-field imbalance directly couples to the far field, forcing the radiation channels to manifest significant circular polarization. As shown in Figs.~6(f) and 6(g), the upward and downward radiations evolve into distinct circular polarization states (indicated by red and blue ellipses). Ultimately, this mechanism drives the drastic splitting of the LCP and RCP transmission spectra shown in Fig.~6(h), revealing the fundamental physical origin of a high CD. These results show that the CD enhancement is not merely a consequence of spectral detuning between two polarization channels. 
Instead, it originates from the symmetry-induced imbalance of the near-field optical chirality. 
The in-plane displacement $\Delta s$ breaks the cancellation between opposite local handedness, thereby converting a nearly achiral resonant mode into a spin-selective resonance with a high far-field CD.

\begin{figure}
    \centering
    \includegraphics[width=1\linewidth]{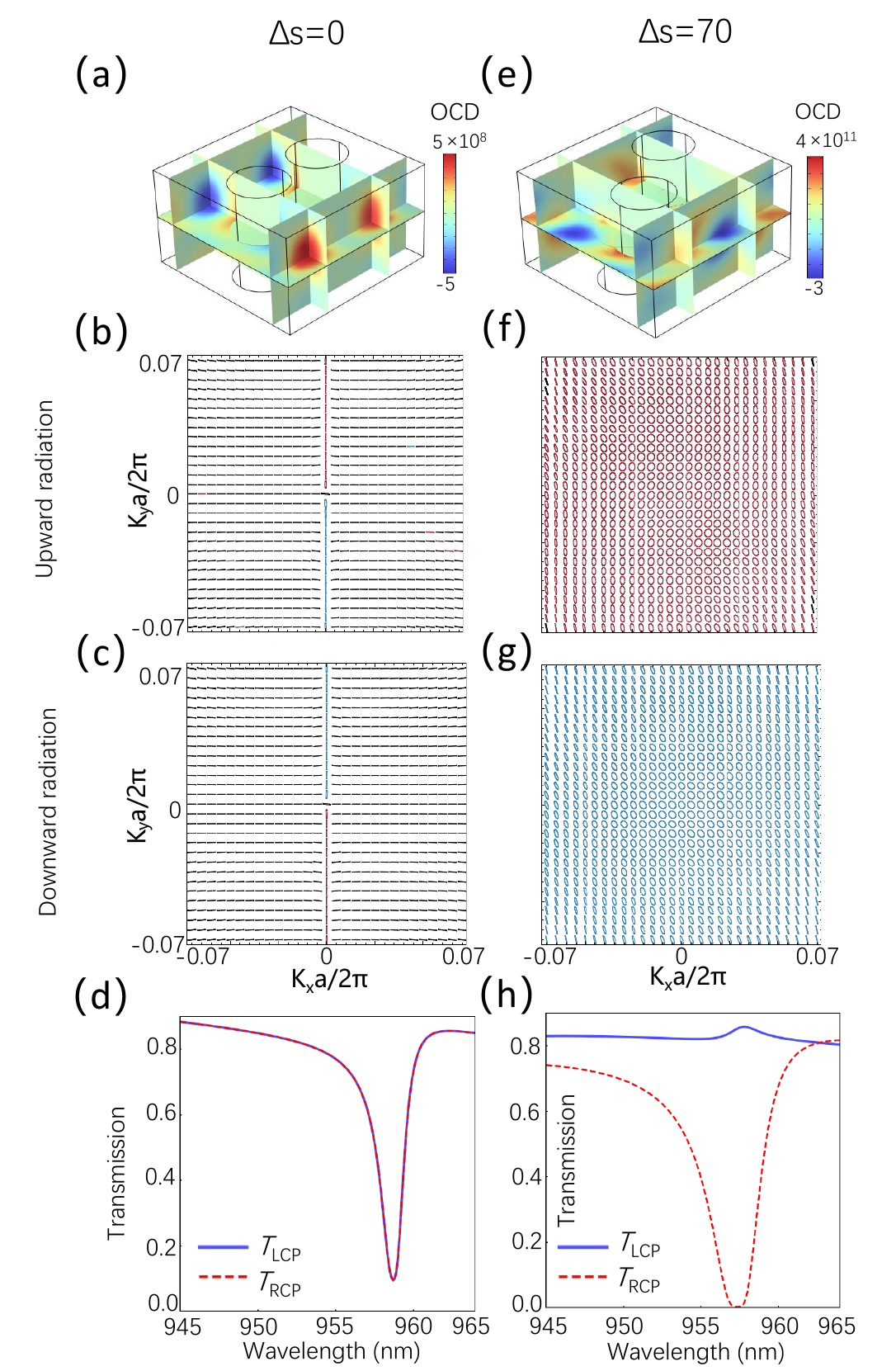}
    \caption{Physical mechanism analysis based on Optical Chirality Density (OCD). The results are organized into two columns: the unperturbed symmetric case ($\Delta s=0$, left column) and the broken-symmetry case ($\Delta s=70$~nm, right column).
\textbf{(a), (e)} Cross-sectional near-field OCD distributions. The symmetric distribution in \textbf{(a)} contrasts with the spatially imbalanced distribution in \textbf{(e)}.
\textbf{(b)-(c), (f)-(g)} Corresponding polarization maps in momentum space for upward (top row of each set) and downward (bottom row of each set) radiation. In the symmetric case, linear polarization is observed for both upward \textbf{(b)} and downward \textbf{(c)} directions. In the broken-symmetry case, the upward \textbf{(f)} and downward \textbf{(g)} radiations evolve into distinct circular polarization states (indicated by red and blue ellipses).
\textbf{(d), (h)} Transmission spectra for LCP (blue solid lines) and RCP (red dashed lines) light. The degenerate spectra in \textbf{(d)} split drastically in \textbf{(h)}, confirming the generation of large CD induced by the symmetry breaking.}
\label{fig:ocd_mechanism}
\end{figure}

\subsection{Active modulation of CD by conductivity tuning}

Finally, we demonstrate that the chiral response of the bilayer PhC slab can be actively modulated by combining a collective in-plane displacement with a tunable material conductivity. Unlike the interlayer displacement $\Delta d$, which produces Janus radiation by lifting the mirror-imposed equivalence between the upward and downward radiation channels, here the same in-plane displacement pattern is applied to both layers. As shown in Fig.~7(a), the two air holes within each unit cell are shifted in opposite directions along the $y$ axis by a distance $\Delta p$, forming an antiphase displacement configuration. Because this perturbation is applied identically to the upper and lower slabs, it does not introduce a relative displacement between the two layers. Instead, it modifies the in-plane modal field distribution and generates a chiral resonant response within a geometrically equivalent bilayer configuration.

We then introduce a tunable bulk conductivity $\sigma$ as a dissipative control parameter, which is assigned only to the dielectric PhC slabs while the substrate and superstrate remain lossless. Here, $\sigma$ is expressed in units of $\mathrm{S/m}$. Although the introduced conductivity is isotropic, the antiphase-displaced geometry produces different near-field overlaps under RCP and LCP illumination. To quantify this effect, we calculate the integrated ohmic loss inside the conductive PhC slabs,~\cite{balanis2012advanced,jackson1999classical}
\begin{equation}
    P_{\mathrm{ohm}}
    =
    \frac{1}{2}
    \int_{\mathrm{slab}}
    \sigma |\mathbf{E}|^2 dV 
    \label{eq:ohmic_loss}
\end{equation}
As shown in Figs.~7(e) and 7(f), the LCP channel exhibits a larger integrated ohmic loss than the RCP channel, and the loss contrast $\Delta P_{\mathrm{ohm}}=P_{\mathrm{ohm}}^{\mathrm{LCP}}-P_{\mathrm{ohm}}^{\mathrm{RCP}}$ increases with $\sigma$. This confirms that the isotropic conductivity produces polarization-dependent damping through the handedness-dependent near-field overlap inside the conductive PhC slabs.

Figures~7(b) and 7(c) show the evolution of the transmission spectra under RCP and LCP illumination, respectively, as $\sigma$ increases. A clear polarization-dependent response is observed. The RCP spectrum remains relatively stable over the considered conductivity range, whereas the resonant feature under LCP excitation is strongly suppressed as $\sigma$ increases. This contrast indicates that the two circular-polarization channels have different sensitivities to conductivity-induced loss. As a result, the transmission contrast between LCP and RCP is progressively reduced with increasing $\sigma$, giving rise to continuous modulation of the CD. As shown in Fig.~7(d), the CD amplitude decreases markedly with increasing conductivity. In the low-conductivity state, the system supports a high CD corresponding to a strongly chiral state. As $\sigma$ increases, the CD rapidly decreases, indicating a transition toward an effectively achiral state. This continuous evolution enables both on--off modulation of chirality and analog tuning of the CD amplitude. Such dynamic control highlights the potential of the proposed design for ultrafast chiral switches, tunable polarization modulators, and reconfigurable photonic circuits.

\begin{figure}
    \centering
    \includegraphics[width=1\linewidth]{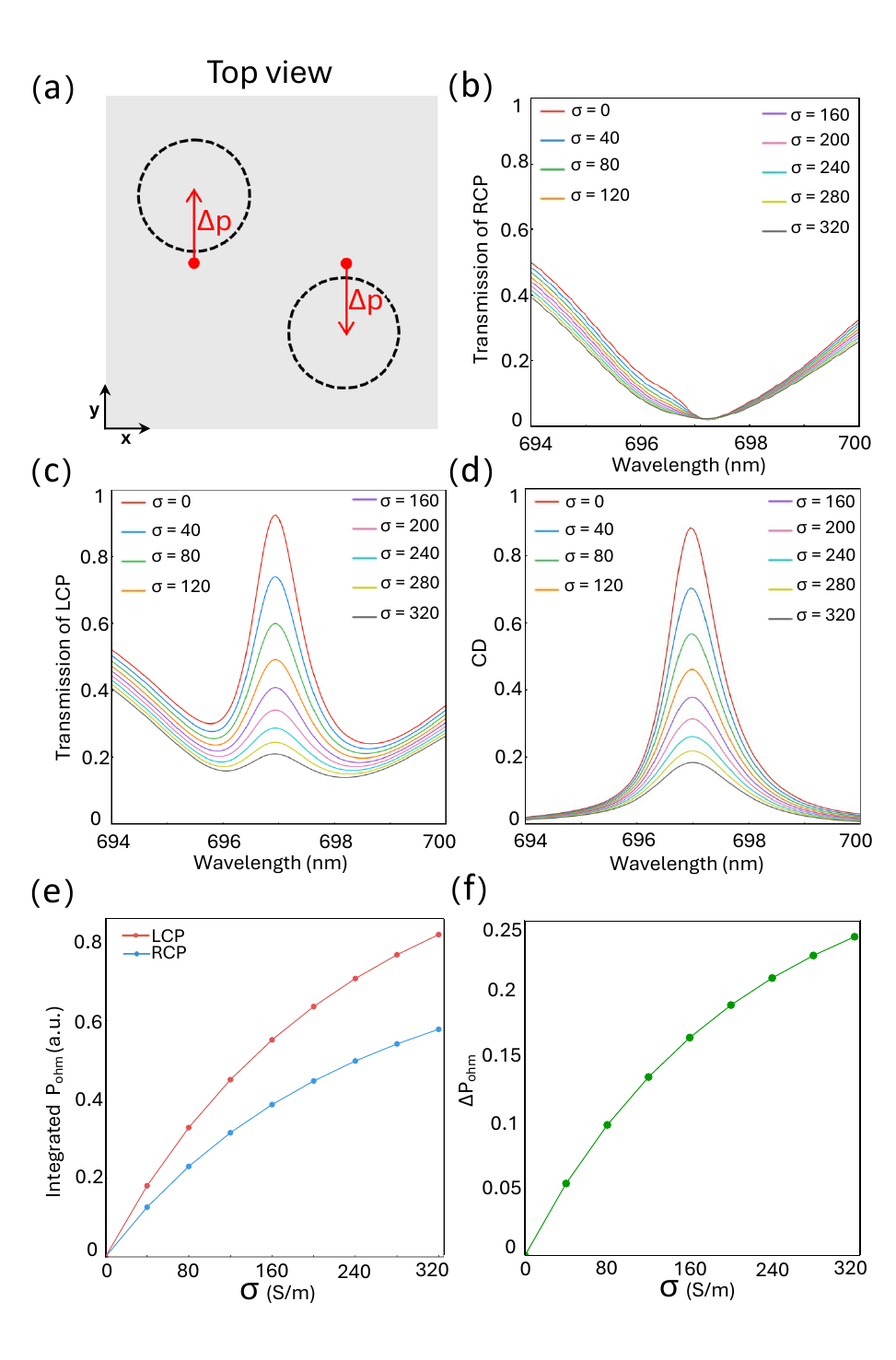}
   \caption{Dynamic modulation of the chiral response by bulk conductivity. 
\textbf{(a)} Top view of the collective antiphase displacement configuration with $\Delta p=80~\mathrm{nm}$, where the two air holes are shifted oppositely along the $y$ axis and the same displacement pattern is applied to both layers. 
\textbf{(b), (c)} Transmission spectra for RCP and LCP incidence, respectively, as a function of the bulk conductivity $\sigma$ assigned to the dielectric PhC slabs, with $\sigma$ in units of $\mathrm{S/m}$. 
\textbf{(d)} CD spectra versus $\sigma$, showing active modulation of the chiral response. 
\textbf{(e)} Integrated ohmic loss $P_{\mathrm{ohm}}$ inside the conductive PhC slabs under RCP and LCP illumination. 
\textbf{(f)} Ohmic-loss contrast $\Delta P_{\mathrm{ohm}}=P_{\mathrm{ohm}}^{\mathrm{LCP}}-P_{\mathrm{ohm}}^{\mathrm{RCP}}$, confirming the polarization-dependent dissipative response.}
\label{fig:active_tuning}
\end{figure}

\begin{figure*}
    \centering
    \includegraphics[width=1\linewidth]{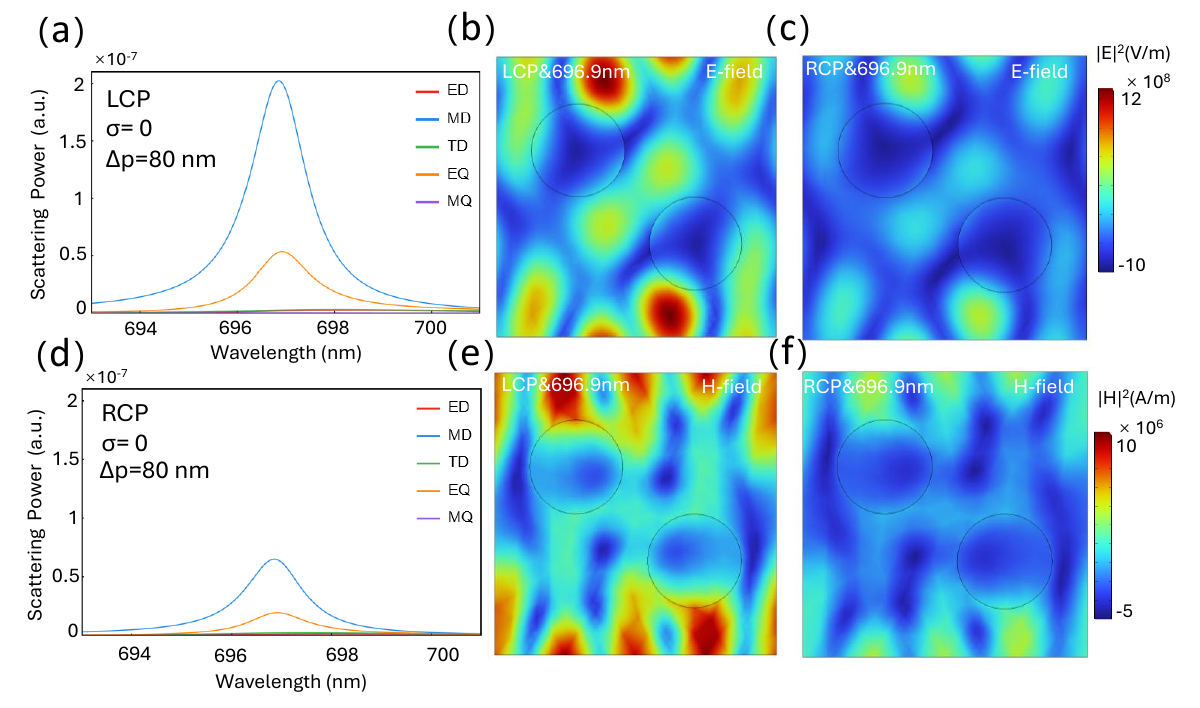}
    \caption{Physical mechanism revealed by multipole decomposition and field distributions. 
\textbf{(a), (d)} Multipole scattering spectra under LCP and RCP incidence at $\Delta p=80$~nm ($\sigma=0$). 
\textbf{(b)-(c)} Electric field ($|E|$) and \textbf{(e)-(f)} magnetic field ($|H|$) distributions in the $x$-$y$ plane at the resonance wavelength ($\lambda=696.9$~nm). The significant enhancement of the MD mode under LCP excitation confirms its strong spin-selective excitation characteristic.}
\label{fig:multipole_mechanism}
\end{figure*}

To elucidate the physical origin of the chiral resonance, we analyze the radiation field using the multipole decomposition method in the Cartesian coordinate system. The total scattering power $I$ can be decomposed into a superposition of contributions from different multipole moments:
\begin{equation}
\begin{split}
    I \cong & \frac{2\omega^4}{3c^3} |\mathbf{P}|^2 + \frac{2\omega^4}{3c^3} |\mathbf{M}|^2 + \frac{2\omega^6}{3c^5} |\mathbf{T}|^2 \\
    & + \frac{\omega^6}{5c^5} \sum_{\alpha,\beta} |Q_{\alpha,\beta}^{(e)}|^2 + \frac{\omega^6}{40c^5} \sum_{\alpha,\beta} |Q_{\alpha,\beta}^{(m)}|^2
\end{split}
\label{eq:multipole_power}
\end{equation}
where $\mathbf{P}$, $\mathbf{M}$, $\mathbf{T}$, $Q^{(e)}$, and $Q^{(m)}$ represent the electric dipole (ED), magnetic dipole (MD), toroidal dipole (TD), electric quadrupole (EQ), and magnetic quadrupole (MQ), respectively. Based on the induced displacement current density $\mathbf{J}(\mathbf{r})$, these multipole moments are explicitly defined as\cite{zeng2025reconfigurable}:
\begin{align}
    \mathbf{P} &= \frac{1}{i\omega} \int \mathbf{J} d^3r \label{eq:ED_def} \\
    \mathbf{M} &= \frac{1}{2c} \int (\mathbf{r} \times \mathbf{J}) d^3r, \label{eq:MD_def} \\
    \mathbf{T} &= \frac{1}{10c} \int [(\mathbf{r} \cdot \mathbf{J})\mathbf{r} - 2r^2\mathbf{J}] d^3r
    \label{eq:TD_def} \\
    Q_{\alpha,\beta}^{(e)} &= \frac{1}{2i\omega} \int [r_\alpha J_\beta + r_\beta J_\alpha - \frac{2}{3}(\mathbf{r} \cdot \mathbf{J})\delta_{\alpha,\beta}] d^3r \label{eq:EQ_def} \\
    Q_{\alpha,\beta}^{(m)} &= \frac{1}{3c} \int [(\mathbf{r} \times \mathbf{J})_\alpha r_\beta + (\mathbf{r} \times \mathbf{J})_\beta r_\alpha] d^3r \label{eq:MQ_def}
\end{align}

where $\omega$ denotes the angular frequency, $c$ is the speed of light, and indices $\alpha, \beta \in \{x, y, z\}$.

Quantitative analysis of the scattering power spectra [Figs.~8(a) and 8(d)] reveals that the magnetic dipole (MD) contribution overwhelmingly dominates at the resonant wavelength ($\lambda = 696.9$~nm), exceeding all other multipole components. This result indicates that the observed chiral resonance stems from the excitation of the MD mode, which is associated with loop-like displacement currents circulating within the structure, as described by the cross-product term in Eq.~\eqref{eq:MD_def}.

More importantly, the MD mode exhibits a pronounced spin-selective coupling characteristic. By comparing Figs.~8(a) and 8(d), it is evident that while LCP light strongly excites the MD resonance, the excitation is drastically suppressed under RCP incidence, resulting in negligible scattering efficiency. This immense disparity reveals the physical mechanism of the chiral response. The structural asymmetry enables a strong coupling between the MD mode and LCP light, whereas the coupling with RCP light is strongly suppressed. This conclusion is further corroborated by the near-field electromagnetic distributions. As shown in Figs.~8(b) and 8(e), LCP excitation induces strong localized electric and magnetic fields within the structure. In sharp contrast, the field intensity under RCP excitation [Figs.~8(c) and 8(f)] is extremely weak. This contrast in near-field energy localization visually verifies the metasurface's capability for the selective trapping and radiation of photons based on their polarization handedness states.

\section{Conclusion}

In summary, we have demonstrated that a bilayer all-dielectric PhC slab provides a unified platform for the controlled realization and active modulation of Janus, Janus--chiral, and chiral quasi-BIC responses through distinct geometric symmetry-breaking pathways and conductivity-based dissipative tuning. An interlayer displacement breaks the out-of-plane mirror symmetry $\sigma_z$, decouples the upward and downward radiation channels, and drives an asymmetric migration of polarization singularities, producing a Janus BIC. A subsequent diagonal in-plane displacement reconstructs the polarization topology around the $\Gamma$ point and yields a Janus-chiral BIC with handedness-selective far-field response, while the $\Gamma$-point BIC condition is preserved throughout. Independent in-plane displacements instead generate chiral quasi-BICs whose CD amplitude and resonance wavelength are continuously tunable, and a collective antiphase displacement combined with finite material conductivity enables active, reversible switching of the chiral response without structural reconfiguration. Near-field optical-chirality and multipole analyses identify a symmetry-induced imbalance of local handedness and a spin-selective magnetic-dipole resonance as the microscopic origins of the chiral far-field response. These results clarify how Janus radiation topology, polarization singularities and intrinsic optical chirality are connected through orthogonal symmetry-breaking channels in bilayer geometries, and show that the divergent $\Gamma$-point $Q$ can be preserved across the Janus-to-chiral evolution. The framework establishes a scalable route to reconfigurable high-$Q$ chiral photonics and is naturally extensible to multilayer architectures and non-Hermitian settings, where gain-loss balance modifies the underlying topological-charge structure.

\begin{acknowledgements}
H. Y. Meng is supported by the Natural Science Foundation of Hunan Province (2023JJ40612). Y. S. Ang acknowledges the supports from the Kwan Im Thong Hood Cho Temple Early Career Chair Professorship and the Singapore Ministry of Education (MOE) Academic Research Fund (AcRF) Tier 2 Grant (MOE-T2EP50224-0021).

\end{acknowledgements}

\section*{Author Declarations}

\subsection*{Conflict of Interest}
\noindent The authors declare that there are no conflicts of interest.

\section*{Data Availability}
The data that support the findings of this study are available from the corresponding author upon reasonable request.

\bibliographystyle{apsrev4-2}
\bibliography{references}

\end{document}